\documentclass[letterpaper,prl,twocolumn,noshowpacs,aps,superscriptaddress,floatfix]{revtex4}

\usepackage{multirow,amssymb,amsbsy,amsmath}
\usepackage{CJK}
\usepackage{graphicx}
\usepackage{dcolumn}
\usepackage{bm}
\usepackage{ifthen}
\usepackage{booktabs}
\usepackage{nicefrac}

\usepackage{float}
\usepackage{color}
\usepackage{dsfont}

\begin{document}

\title{Experimental investigation of direct non-Hermitian measurement and uncertainty relation towards high-dimensional quantum domain}

\author{Yi-Tao Wang}
\thanks{These authors contributed equally to this work.}
\affiliation{CAS Key Laboratory of Quantum Information, University of Science and Technology of China, Hefei, Anhui 230026, China}
\affiliation{Anhui Province Key Laboratory of Quantum Network, University of Science and Technology of China, Hefei, Anhui 230026, China}
\affiliation{CAS Center For Excellence in Quantum Information and Quantum Physics, University of Science and Technology of China, Hefei, Anhui 230026, China}

\author{Zhao-An Wang}
\thanks{These authors contributed equally to this work.}
\affiliation{CAS Key Laboratory of Quantum Information, University of Science and Technology of China, Hefei, Anhui 230026, China}
\affiliation{Anhui Province Key Laboratory of Quantum Network, University of Science and Technology of China, Hefei, Anhui 230026, China}
\affiliation{CAS Center For Excellence in Quantum Information and Quantum Physics, University of Science and Technology of China, Hefei, Anhui 230026, China}

\author{Zhi-Peng Li}
\affiliation{CAS Key Laboratory of Quantum Information, University of Science and Technology of China, Hefei, Anhui 230026, China}
\affiliation{Anhui Province Key Laboratory of Quantum Network, University of Science and Technology of China, Hefei, Anhui 230026, China}
\affiliation{CAS Center For Excellence in Quantum Information and Quantum Physics, University of Science and Technology of China, Hefei, Anhui 230026, China}

\author{Xiao-Dong Zeng}
\affiliation{CAS Key Laboratory of Quantum Information, University of Science and Technology of China, Hefei, Anhui 230026, China}
\affiliation{Anhui Province Key Laboratory of Quantum Network, University of Science and Technology of China, Hefei, Anhui 230026, China}
\affiliation{CAS Center For Excellence in Quantum Information and Quantum Physics, University of Science and Technology of China, Hefei, Anhui 230026, China}

\author{Jia-Ming Ren}
\affiliation{CAS Key Laboratory of Quantum Information, University of Science and Technology of China, Hefei, Anhui 230026, China}
\affiliation{Anhui Province Key Laboratory of Quantum Network, University of Science and Technology of China, Hefei, Anhui 230026, China}
\affiliation{CAS Center For Excellence in Quantum Information and Quantum Physics, University of Science and Technology of China, Hefei, Anhui 230026, China}

\author{Wei Liu}
\affiliation{CAS Key Laboratory of Quantum Information, University of Science and Technology of China, Hefei, Anhui 230026, China}
\affiliation{Anhui Province Key Laboratory of Quantum Network, University of Science and Technology of China, Hefei, Anhui 230026, China}
\affiliation{CAS Center For Excellence in Quantum Information and Quantum Physics, University of Science and Technology of China, Hefei, Anhui 230026, China}

\author{Yuan-Ze Yang}
\affiliation{CAS Key Laboratory of Quantum Information, University of Science and Technology of China, Hefei, Anhui 230026, China}
\affiliation{Anhui Province Key Laboratory of Quantum Network, University of Science and Technology of China, Hefei, Anhui 230026, China}
\affiliation{CAS Center For Excellence in Quantum Information and Quantum Physics, University of Science and Technology of China, Hefei, Anhui 230026, China}

\author{Nai-Jie Guo}
\affiliation{CAS Key Laboratory of Quantum Information, University of Science and Technology of China, Hefei, Anhui 230026, China}
\affiliation{Anhui Province Key Laboratory of Quantum Network, University of Science and Technology of China, Hefei, Anhui 230026, China}
\affiliation{CAS Center For Excellence in Quantum Information and Quantum Physics, University of Science and Technology of China, Hefei, Anhui 230026, China}
\affiliation{Hefei National Laboratory, University of Science and Technology of China, Hefei, Anhui 230088, China}

\author{Lin-Ke Xie}
\affiliation{CAS Key Laboratory of Quantum Information, University of Science and Technology of China, Hefei, Anhui 230026, China}
\affiliation{Anhui Province Key Laboratory of Quantum Network, University of Science and Technology of China, Hefei, Anhui 230026, China}
\affiliation{CAS Center For Excellence in Quantum Information and Quantum Physics, University of Science and Technology of China, Hefei, Anhui 230026, China}

\author{Jun-You Liu}
\affiliation{CAS Key Laboratory of Quantum Information, University of Science and Technology of China, Hefei, Anhui 230026, China}
\affiliation{Anhui Province Key Laboratory of Quantum Network, University of Science and Technology of China, Hefei, Anhui 230026, China}
\affiliation{CAS Center For Excellence in Quantum Information and Quantum Physics, University of Science and Technology of China, Hefei, Anhui 230026, China}
\affiliation{Hefei National Laboratory, University of Science and Technology of China, Hefei, Anhui 230088, China}

\author{Yu-Hang Ma}
\affiliation{CAS Key Laboratory of Quantum Information, University of Science and Technology of China, Hefei, Anhui 230026, China}
\affiliation{Anhui Province Key Laboratory of Quantum Network, University of Science and Technology of China, Hefei, Anhui 230026, China}
\affiliation{CAS Center For Excellence in Quantum Information and Quantum Physics, University of Science and Technology of China, Hefei, Anhui 230026, China}

\author{Jian-Shun Tang}
\email{tjs@ustc.edu.cn}
\affiliation{CAS Key Laboratory of Quantum Information, University of Science and Technology of China, Hefei, Anhui 230026, China}
\affiliation{Anhui Province Key Laboratory of Quantum Network, University of Science and Technology of China, Hefei, Anhui 230026, China}
\affiliation{CAS Center For Excellence in Quantum Information and Quantum Physics, University of Science and Technology of China, Hefei, Anhui 230026, China}
\affiliation{Hefei National Laboratory, University of Science and Technology of China, Hefei, Anhui 230088, China}

\author{Chengjie Zhang}
\email{chengjie.zhang@gmail.com}
\affiliation{School of Physical Science and Technology, Ningbo University, Ningbo 315211, China}
\affiliation{Hefei National Laboratory, University of Science and Technology of China, Hefei, Anhui 230088, China}

\author{Chuan-Feng Li}
\email{cfli@ustc.edu.cn}
\affiliation{CAS Key Laboratory of Quantum Information, University of Science and Technology of China, Hefei, Anhui 230026, China}
\affiliation{Anhui Province Key Laboratory of Quantum Network, University of Science and Technology of China, Hefei, Anhui 230026, China}
\affiliation{CAS Center For Excellence in Quantum Information and Quantum Physics, University of Science and Technology of China, Hefei, Anhui 230026, China}
\affiliation{Hefei National Laboratory, University of Science and Technology of China, Hefei, Anhui 230088, China}

\author{Guang-Can Guo}
\affiliation{CAS Key Laboratory of Quantum Information, University of Science and Technology of China, Hefei, Anhui 230026, China}
\affiliation{Anhui Province Key Laboratory of Quantum Network, University of Science and Technology of China, Hefei, Anhui 230026, China}
\affiliation{CAS Center For Excellence in Quantum Information and Quantum Physics, University of Science and Technology of China, Hefei, Anhui 230026, China}
\affiliation{Hefei National Laboratory, University of Science and Technology of China, Hefei, Anhui 230088, China}

\date{\today}

\begin{abstract}
Non-Hermitian dynamics in quantum systems have unveiled novel phenomena, yet the implementation of valid non-Hermitian quantum measurement remains a challenge, because a universal quantum projective mechanism on the complete but skewed non-Hermitian eigenstates is not explicit in experiment. This limitation hinders the direct acquisition of non-Hermitian observable statistics (e.g., non-Hermitian population dynamics), also constrains investigations of non-Hermitian quantum measurement properties such as uncertainty relation. Here, we address these challenges by presenting a non-Hermitian projective protocol and investigating the non-Hermitian uncertainty relation. We derive the uncertainty relation for pseudo-Hermitian (PH) observables that is generalized beyond the Hermitian ones. We then investigate the projective properties of general quantum states onto complete non-Hermitian eigenvectors, and present a quantum simulating method to apply the valid non-Hermitian projective measurement on a direct-sum dilated space. Subsequently, we experimentally construct a quantum simulator in the quantum optical circuit and realize the 3-dimensional non-Hermitian quantum measurement on the single-photon qutrit. Employing this platform, we explore the uncertainty relation experimentally with different PH metrics. Our non-Hermitian quantum measurement method is state-independent and outputs directly the non-Hermitian quantum projective statistics, paving the way for studies of extensive non-Hermitian observable in quantum domain.
\end{abstract}

\maketitle

\section*{Introduction}
Quantum measurement can be interpreted physically as a projective process by von Neumann measurement, where a quantum state is projected onto independent eigenvectors with observed eigenvalues \cite{Nielsen2010}. Typically, physical observables are represented by Hermitian operators due to their possession of complete orthogonal eigenvectors and real eigenvalues. However, non-Hermitian Hamiltonians may also possess real eigenvalues with the complete skewed eigenstates, and non-Hermitian dynamics involving with novel phenomena have been observed in various physical systems recently \cite{Bender1998,Ozdemir2019,Miri2019,Wang2020,Ding2022}. These non-Hermitian eigenstates exhibit independent and stable dynamical evolution in the non-Hermitian systems. This non-Hermitian nature indicates that the effective projective measurement is required to acquire the evolution state of non-Hermitian system, such as the dynamical population on eigenstates. Crucially, it is necessary to explore that how a quantum state is projected on the nonorthogonal but independent eigenstates (as depicted in Fig. 1(a)) in measurement.

\begin{figure}[tb]
\includegraphics[width=2.8in]{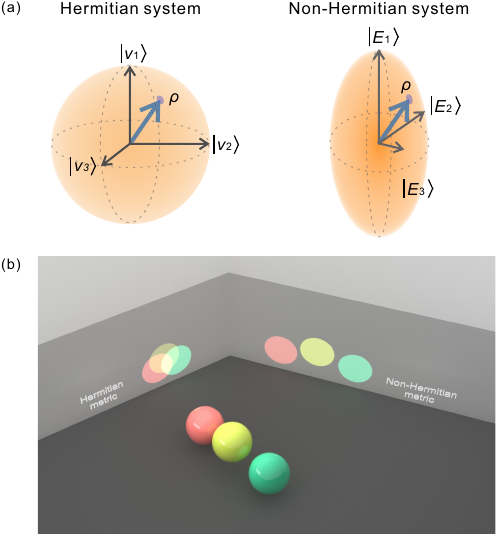}
\renewcommand{\thefigure}{1}
\caption{Non-Hermitian eigenvectors and observable. (a) A quantum state $\rho$ located in 3-dimensional Hermitian (left) and non-Hermitian (right) systems. The Hermitian eigenvectors $|v_k\rangle$ $(k=1,2,3)$ satisfy standard orthonormal condition, while the non-Hermitian eigenvectors  $|E_k\rangle$ $(k=1,2,3)$ typically do not. (b) Non-Hermitian observable manifesting independent eigenmodes (depicted as three balls) in non-Hermitian dynamics. The non-Hermitian observable exhibits the overlapping projection under the Hermitian metric, but can be separable as the orthogonal projection under the non-Hermitian inner product metric.}
\center
\end{figure}

In particular, a class of non-Hermitian operators known as pseudo-Hermitian (PH) operators may exhibit real eigenvalue spectra and possess the complete orthogonal eigenvectors under the PH inner product \cite{Mostafazadeh2002,Mostafazadeh20022,Mostafazadeh2007,Supp}. These PH operators represent typical non-Hermitian physical observables, and the PH metric provides an effective projective mechanism onto independent but skewed non-Hermitian eigenstates, as depicted in Fig. 1(b). Furthermore, typical quantum measurement properties, such as the uncertainty relation, is required to be studied for PH observable. The uncertainty relation reveals the fundamental quantum nature, i.e., the inherent measurement uncertainty stemming from observable noncommutation \cite{Heisenberg1927,Robertson1929,Maccone2014,Ma2017,Maassen1988}, and plays essential roles in various quantum properties and applications associated with quantum measurement \cite{Coles2017,Busch2014,Guhne2004,Hou2021}. While the common uncertainty relations concentrate mainly on Hermitian observables under the Dirac inner product, PH observables and non-Hermitian measurement hold the variable inner product metric. Therefore, studying the PH uncertainty relation under the PH metric is significant to explore the quantum measurement properties of non-Hermitian systems.

However, the realization of general non-Hermitian quantum measurement remains a challenge in experiment, primarily due to the absence of a state-independent projective protocol on complete but skewed non-Hermitian eigenvectors. Such a protocol is crucial for the developing a feasible experimental method for universal non-Hermitian measurement. Biorthogonal decomposition was utilized to derive the population in various non-Hermitian dynamics \cite{Ding2022,Brody2014}, but this derivation relies on acquiring the density matrix of states, meanwhile, the direct non-Hermitian quantum measurement without state tomography is unattainable. Furthermore, while non-Hermitian quantum dynamics have been realized by diverse methods \cite{Miri2019,Wang2020,Xiao2019}, but mainly concentrates on the evolution rather than measurement. In contrast, positive operator value measure (POVM) is a typical generalized measurement without orthogonality restriction, but the POVM can be dilated to standard orthogonal measurement by Naimark's theorem and the projected quantum state in POVM is uncertain \cite{Nielsen2010}. Given that non-Hermitian measurement exhibits distinct mechanisms from POVM (see \emph{Discussions}), a valid non-Hermitian projective measurement method remains to be explored.

Here, we investigate the general projective properties of PH observable for the arbitrary quantum state, and then develop a quantum simulation method to realize the non-Hermitian projective measurement, which employs direct-sum dilated projection combining with the postselection. Next, we construct experimentally a quantum photonic simulator, and carry out the 3-dimensional non-Hermitian quantum measurement on single-photon qutrits using different PH metrics. Notably, we directly acquire measured results from the output statistical distribution without the need of state tomography. Furthermore, we derive the PH uncertainty relation with positive definite PH metric and investigate it experimentally within the quantum simulator. Our experimental results demonstrate that the PH uncertainty relation indeed holds under the positive definite PH metric, but could be violated under the non-positive definite PH metric. Overall, our work provides and demonstrates an effective non-Hermitian method for directly measuring non-Hermitian quantum systems.

\section*{Results}
\textbf{Pseudo-Hermitian observable and uncertainty relation.} A PH operator $H$ satisfies the condition $H^{\dag}=\eta H \eta^{-1}$, where $\eta$ is an invertible Hermitian matrix representing the PH metric. PH metric redefines the inner product as $\langle\psi_1|\psi_2\rangle_{\eta}:=\langle\psi_1|\eta|\psi_2\rangle$ where $|\psi_1\rangle$ and $|\psi_2\rangle$ are two arbitrary pure quantum states. The general PH operator typically possesses real eigenvalues or complex eigenvalues in conjugated pairs. In particular, for an $n$-dimensional PH operator $H$ with eigenstates $\{|E_k\rangle\}$ and eigenvalues $\{e_k\}$ such that $H|E_k\rangle=e_k|E_k\rangle$ ($k=1,\cdot\cdot\cdot,n$), the following spectral properties hold under the conditions of non-degeneration ($e_k\neq e_l$ when $k\neq l$ ) and non-vanishing $\eta$-norm ($\langle E_k|\eta|E_k\rangle\neq 0$ for all $k$): (1) all eigenvalues $\{e_k\}$ are real values, (2) the $\eta$-normalized eigenvectors $\{|E_k\rangle\}$ satisfy $\eta$-orthogonality $\langle E_k|\eta|E_l\rangle=\pm\delta_{kl}$ and completeness $\sum_k \frac{|E_k\rangle\langle E_k|\eta}{\langle E_k|\eta|E_k\rangle}=\mathds{1}$ \cite{Supp}. These properties reveal that the eigenspectrum of a PH operator exhibits observable characters with real observed values under the PH metric, furthermore, can be viewed as a natural generalization from Hermitian observables, as the PH operator and inner product reduce to the normal Hermitian ones when $\eta = \mathds{1}$.

\begin{figure*}[tb]
\includegraphics[width=7in]{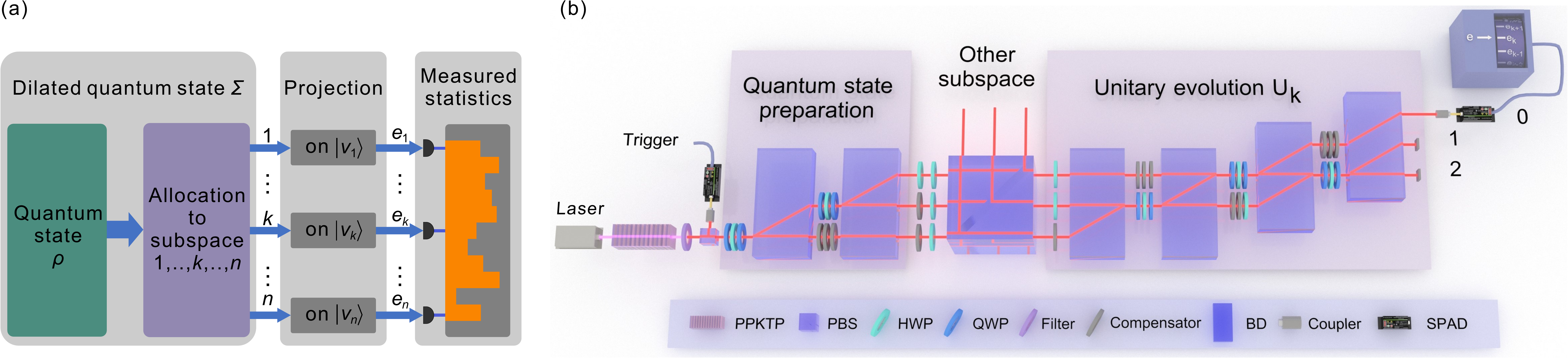}
\renewcommand{\thefigure}{2}
\caption{Non-Hermitian quantum measurement simulator. (a) State-independent non-Hermitian projective scheme. A quantum state $\rho$ is input to the quantum simulator for an $n$-dimensional non-Hermitian projective measurement. The quantum state is then allocated to the subspaces 1 to $n$ and projected on the corresponding $\{|v_k\rangle\}$ ($k=1,...,n$), where the allocating proportion and $\{|v_k\rangle\}$ are determined by the PH observable. Subsequently, the projective events are detected and labeled by the eigenvalues $\{|e_k\rangle\}$ ($k=1,...,n$), and the measurement results are obtained by the statistical distribution of detected events. (b) Experimental setup for one subspace. The 3-dimensional quantum state is prepared on the path state of the single photon in the quantum state preparation module. The single photon is then allocated to the subspace $k$ ($k=1,...,n$) after transmitting the polarizing beam splitter (PBS), where the allocating proportion is regulated by the single-photon polarization. Next, the single photon undergoes the 3-dimensional unitary evolution $U_k$ and is projected onto $|v_k\rangle$ at the output path 0. The projected photon is detected by the single-photon avalanche diode (SPAD) and labeled as the projective event with the eigenvalue $e_k$. BD, beam displacer. HWP, half-wave plate. QWP, quarter-wave plate.}
\center
\end{figure*}

According to the completeness of PH eigenvectors, the $\eta$-trace $\text{Tr}[\bullet]=\sum_k\frac{\langle E_k|\eta \bullet|E_k\rangle}{\langle E_k|\eta|E_k\rangle}$ is equivalent to the standard trace $\text{Tr}[\bullet]=\sum_k\langle\psi_k|\bullet|\psi_k\rangle$ where $\{|\psi_k\rangle\}$ represents the standard orthonormal bases. Both the traces are independent of the specific basis set \cite{Supp}. The expecting value and variance of the observable $H$ under the PH inner product can be defined as
\begin{equation}\label{eqdef}
\begin{aligned}
\langle H\rangle_{\eta}&=\text{Tr}[\rho\eta H], \\
\langle (\Delta H)^2\rangle_{\eta}&=\text{Tr}[\rho\eta (H-\langle H\rangle_{\eta})^2],
\end{aligned}
\end{equation}
where $\rho$ is the Hermitian density matrix of a quantum state $\eta$-normalized by $\rho\rightarrow \rho/\text{Tr}[\rho\eta]$. Any arbitrary quantum state $\rho$ can be decomposed onto the linearly independent PH eigenvectors as $\rho=\sum_{kl}p_{kl}|E_k\rangle\langle E_l|$, where $p_{kl}=\frac{\langle E_k|\eta\rho\eta |E_l\rangle}{\langle E_k|\eta|E_k\rangle\langle E_l|\eta|E_l\rangle}$ represents the complex coefficient. Utilizing this decomposition, the expecting value and variance of $H$ for $\rho$ can be further derived as

\begin{equation}\label{eqob}
\begin{aligned}
\langle H\rangle_{\eta}&=\sum_{k}p_{kk}\langle E_k|\eta|E_k\rangle e_k,\\
\langle(\Delta H)^2\rangle_{\eta}&=\sum_{k}p_{kk}\langle E_k|\eta|E_k\rangle(e_k-\langle H\rangle_{\eta})^2,
\end{aligned}
\end{equation}
where the coefficient $\langle E_k|\eta|E_k\rangle=\pm 1$ since $\eta$ is not restricted as the positive definite operator. According to Eq. \ref{eqob}, the measurement of PH observable can be interpreted as the projective measurement on the complete $\eta$-orthogonal eigenvectors $\{|E_k\rangle\}$. The projective probabilities on the PH eigenstates are proportional to $\{p_{kk}\}$ involving with the weight coefficient $\langle E_k|\eta|E_k\rangle$. All the $\{p_{kk}\}$ are real values considering $\rho$ is a Hermitian operator, hence the expecting value $\langle H\rangle_{\eta}$ is also a real value.

In the following, we derive the uncertainty relation of two PH operators $A$ and $B$ with the positive definite PH metric $\eta$. Consider a matrix $M$ defined as
\begin{equation}\label{eqM}
\begin{aligned}
M=\left(
\begin{array}{cc}
\langle(\Delta A)^2\rangle_{\eta} & \langle AB\rangle_{\eta}-\langle A\rangle_{\eta}\langle B\rangle_{\eta} \\
 \langle BA\rangle_{\eta}-\langle B\rangle_{\eta}\langle A\rangle_{\eta} & \langle(\Delta B)^2\rangle_{\eta}
\end{array}
\right).
\end{aligned}
\end{equation}
It can be obtained that $\langle A\rangle_{\eta}\geq0$, $\langle B\rangle_{\eta}\geq0$ and $\langle d|M|d\rangle=\langle C^{\dag}\eta C\rangle$, where the vector $|d\rangle=(a,b)^{T}$ with $a$ and $b$ being complex numbers, and the operator $C=a\Delta A+b\Delta B$. Since $\eta$ is positive definite, $\langle C^{\dag}\eta C\rangle$ is always a non-negative number for arbitrary $|d\rangle$. This indicates that $M$ is a positive semi-definite operator. Thus, we can obtain that $\text{det}(M)\geq0$ with $\text{det}(M)=\langle(\Delta A)^2\rangle_{\eta}\langle(\Delta B)^2\rangle_{\eta}-|\langle AB\rangle-\langle A\rangle_{\eta}\langle B\rangle_{\eta}|$, where $(\langle BA\rangle_{\eta}-\langle B\rangle_{\eta}\langle A\rangle_{\eta})^*=\langle AB\rangle_{\eta}-\langle A\rangle_{\eta}\langle B\rangle_{\eta}$ is used. The uncertainty relation for the PH observable $A$ and $B$ under the positive definite PH metric is derived out as
\begin{equation}\label{equn}
\begin{aligned}
\langle(\Delta A)^2\rangle_{\eta}\langle(\Delta B)^2\rangle_{\eta}\geq|\langle AB\rangle_{\eta}-\langle A\rangle_{\eta}\langle B\rangle_{\eta}|^2.
\end{aligned}
\end{equation}
When $A$ and $B$ are commutated, they share common eigenvectors, and thus $\langle AB\rangle_{\eta}=\langle A\rangle_{\eta}\langle B\rangle_{\eta}$. This indicates that the observable non-commutation also influences the uncertainty of PH measurement.

\textbf{Non-Hermitian quantum measurement simulator.} Given that the eigenvectors $\{|E_k\rangle\}$ are $\eta$-orthogonal but generally not standard orthonormal, a state independent projective method on the complete skewed eigenbases is essential for implementing the non-Hermitian measurement of PH observable. Here, we employ the direct-sum dilated orthonormal projection combining with postselection to simulate non-Hermitian quantum measurement. The dilated projective operator is constructed in an $n^2$-dimensional space as $P=P_1\bigoplus...\bigoplus P_n$ that satisfies $P^2=P$. In each $n$-dimensional subspace, the projective operator $P_l=|v_l\rangle\langle v_l|$, where $\langle v_l|v_l\rangle=1$ and $|v_l\rangle$ is orthogonal to all the eigenvectors in $\{|E_k\rangle\}$ except $|E_l\rangle$. The quantum state is dilated as $\Sigma=\frac{1}{N}(\rho_1\bigoplus...\bigoplus\rho_n)$, where $\rho_k=\frac{1}{|\langle v_k|E_k\rangle|^2}\rho$ and the normalizing coefficient $N=\sum_{k} \frac{1}{n|\langle v_k|E_k\rangle|^2}$. In this direct-sum dilation configuration, the projective operation can be applied respectively in each subspace as $\text{Tr}[P\Sigma]=\frac{1}{N}(\text{Tr}[P_1\rho_1]\bigoplus...\bigoplus\text{Tr}[P_n\rho_n])$, and the projective probability in each subspace is proportional to $\text{Tr}[P_k\rho_k]=p_{kk}$ \cite{note1}. Thus, the non-Hermitian projective measurement can be realized by applying the standard orthonormal postselected projection $P$ on the dilated quantum state $\Sigma$, where the projection in the $k$-th subspace corresponds to the projection onto the eigenstate $|E_k\rangle$ of $H$ in the non-Hermitian measurement.

Based on the dilated projective method, we experimentally construct a quantum simulator in the quantum optical circuit, as shown in Fig. 2. The state independent non-Hermitian projective measurement scheme is illustrated in Fig. 2(a), and the experimental setup of the 3-dimensional non-Hermitian measurement for one subspace is shown in Fig. 2(b). The preparation of the dilated quantum state $\Sigma$ involves by two steps: (1) generating heralded single photons via spontaneous parametric down-conversion and preparing the quantum state $\rho$ on the 3-dimensional path state of single photons \cite{Supp}; (2) Regulating the allocation of single photons, which is proportional to $\frac{1}{|\langle v_k|E_k\rangle|^2}$ in each subspace, and then sending the single photons at state $\frac{\rho_k}{|\langle v_k|E_k\rangle|^2}$ into the subsequent subspace projection module. The allocation and projection can be independently applied for each subspace, as the direct-sum subspaces are divisible in our dilation configuration.

In the projection module, we utilize multiple beam displacers and wave plates are utilized to apply the universal 3-dimensional unitary evolution $U_k$ on the path state of the single photon, which satisfies $U_k|v_k\rangle=|0\rangle$ and $|0\rangle$ corresponds to the output path 0 marked in Fig. 2(b) \cite{Supp}. This ensures that the single photon output from $|0\rangle$ is projected onto $|v_k\rangle$. The projected single photon is then postselected on $|0\rangle$ and detected by the single photon avalanche diode (SPAD), with the detected events are recorded within a fixed time window. After regulating the allocation and $U_k$ for every subspace, we can obtain the statistical distribution of the projection onto the complete PH eigenvectors $\{|E_k\rangle\}$ from all the detected events. By labeling the eigenvalues as the observed values and taking $\langle E_k|\eta|E_k\rangle=\pm 1$ as the statistical weight of the detected events in every subspace \cite{note2}, we can obtain the expecting value and variance of the PH observable $H$ from the output statistical distribution of the quantum simulator.

In this non-Hermitian quantum simulator, the quantum state to be measured is input directly, and the measured result is output directly by the detected statistics without the need to derive it from the density matrix of quantum state. Moreover, the projective operator $P$ and the allocation proportion are uniquely determined by the PH observable $H$ but are independent of the quantum state $\rho$. Therefore, our non-Hermitian quantum simulator, based on the dilated projecting method, provide a state-independent realization protocol for non-Hermitian projective measurement.

\textbf{Experimental measurement of PH observable and uncertainty relation.}
Based on the constructed PH quantum simulator, we measure multiple non-Hermitian observables with various quantum states and experimentally investigate the PH uncertainty relation under different PH metrics. The quantum state is set as $|\psi\rangle=\frac{1}{M}(\cos\theta_1\sin\theta_2, \cos\theta_1\cos\theta_2, \sin\theta_1)^{\text{T}}$, where $\theta_1$ and $\theta_2$ are variable parameters and $M$ is the $\eta$-normalizing coefficient that ensures $\text{Tr}[|\psi\rangle\langle\psi|\eta]=1$. We firstly investigate the PH observable with the positive definite PH metric $\eta=\text{diag}(1,1,0.6)$. Two PH observable $A$ and $B$ are measured as following
\begin{equation}\label{eqpab}
\begin{aligned}
A=\left(
\begin{array}{ccc}
0 & 0.3 & 1.2 \\
0.3 & 0 & 0 \\
2 & 0 & 0 \\
\end{array}
\right),
\quad
B=\left(
\begin{array}{ccc}
0 & 2 & -0.6 \\
2 & 0 & 0 \\
-1 & 0 & 0 \\
\end{array}
\right).
\end{aligned}
\end{equation}
The expecting value and variance of $A$ and $B$, for the quantum state with $\theta_1=0$, $\frac{\pi}{2.5}$ and varying $\theta_2$, are shown in Fig. 3. The experimental and theocratical results are consistent with each other, where the theocratical results are derived from Eq. (\ref{eqdef}). This demonstrates the availability of our non-Hermitian quantum simulator for the measurement of PH observable.

\begin{figure}[tb]
\includegraphics[width=3.3in]{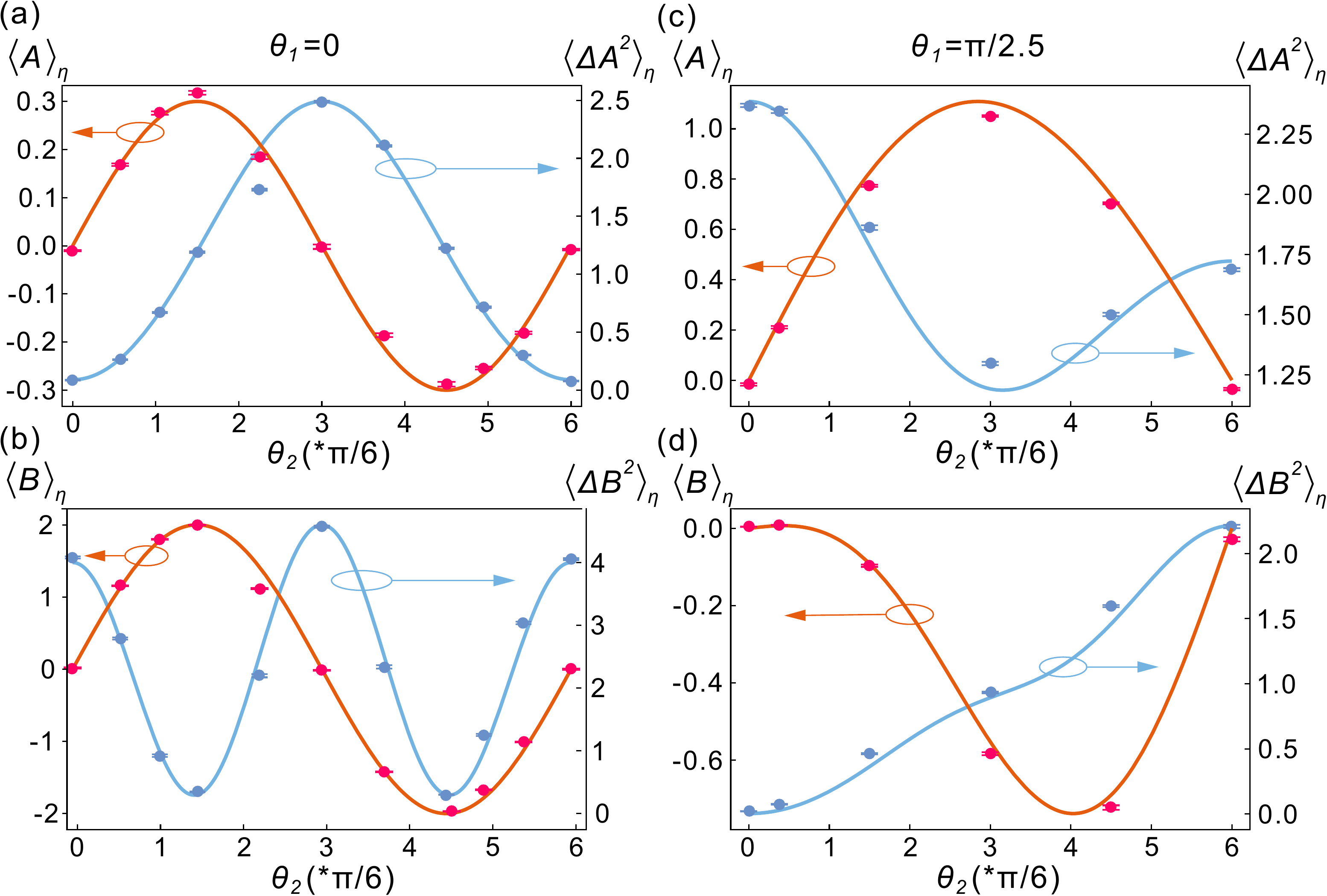}
\renewcommand{\thefigure}{3}
\caption{Results of non-Hermitian projective measurement. The quantum state is $|\psi\rangle=(\cos\theta_1\sin\theta_2, \cos\theta_1\cos\theta_2, \sin\theta_1)^{\text{T}}$ (without normalization) where $\theta_2$ is variable and $\theta_1$ is set as $0$ (for (a) and (b)) and $\frac{\pi}{2.5}$ (for (c) and (d)). The lines are the theoretical results and the dots with error bars are the experimental results. The expecting value and variance of two pseudo-Hermitian observables $A$ and $B$ are displayed. The red lines and dots indicate the expecting value with labels on the left axes. The blue lines and dots indicate the variance with labels on the right axes.}
\center
\end{figure}

Next, we investigate the non-Hermitian uncertainty relation under the positive definite PH metric. Considering the operator $AB$ may not be a PH observable, we take an unbiased estimation by decomposing into two PH observables as $\langle AB\rangle_{\eta}=(1+i)\langle C_1\rangle_{\eta}+(1-i) \langle C_2\rangle_{\eta}$ and measure the PH observables $C_1$ and $C_2$ in the quantum simulator. The experimental results of the uncertainty relation are given by $R=\frac{\langle(\Delta A)^2\rangle_{\eta}\langle(\Delta B)^2\rangle_{\eta}}{|\langle AB\rangle_{\eta}-\langle A\rangle_{\eta}\langle B\rangle_{\eta}|^2}$ in Fig. 4(a) and 4(b). For the quantum states with $\theta_1=0$, $\frac{\pi}{2.5}$ and varying $\theta_2$, all the observed results exhibit $R\geq1$. It demonstrates that the uncertainty relation can hold under the positive definite PH metric for various quantum states.

Finally, we investigate the uncertainty relation under the non-positive definite PH metric $\eta=\text{diag}(1,1,-1)$. Two PH observables to be measured are taken as
\begin{equation}\label{eqnab}
\begin{aligned}
A=\left(
\begin{array}{ccc}
0 & 2 & -1 \\
2 & 0 & 0 \\
1 & 0 & 0 \\
\end{array}
\right),
\quad
B=\left(
\begin{array}{ccc}
0 & 4 & -3 \\
4 & 0 & 0 \\
3 & 0 & 0 \\
\end{array}
\right).
\end{aligned}
\end{equation}
The uncertainty relation result $R$ for various quantum states is shown in Fig. 4(c) and 4(d). It exhibits $R\geq1$ in Fig. 4(d) for the quantum states with $\theta_1=\frac{\pi}{2.5}$, but $R\leq 1$ with $\theta_1=0$. It demonstrates that the uncertainty relation (\ref{equn}) may be violated under the non-positive definite PH metric. Particularly, the variance $\langle(\Delta A)^2\rangle_{\eta}$ and $\langle(\Delta B)^2\rangle_{\eta}$ could be negative with non-positive definite $\eta$, which leads to the negative $R$ in Fig. 4(c).

\begin{figure}[tb]
\includegraphics[width=3.3in]{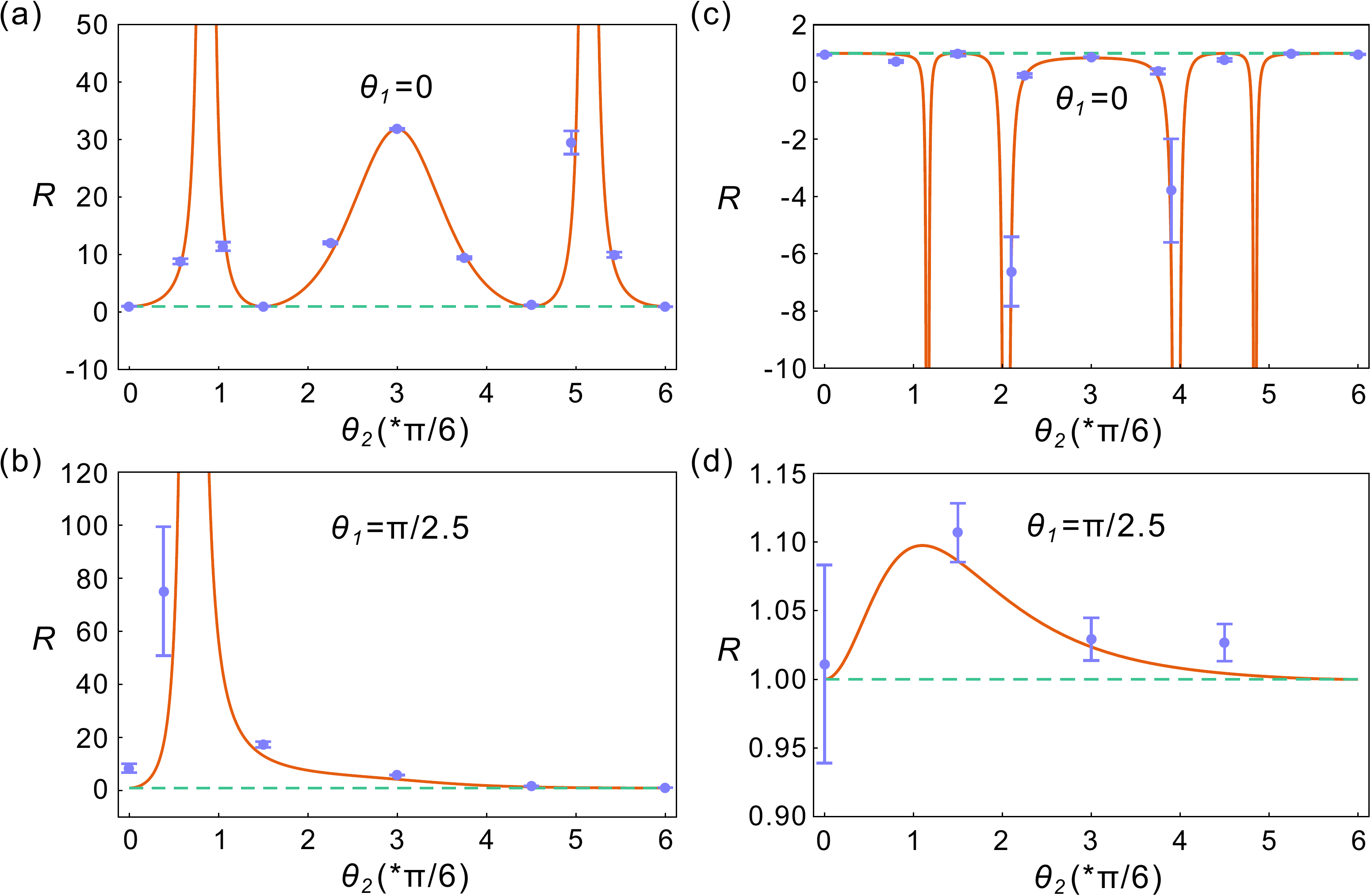}
\renewcommand{\thefigure}{4}
\caption{Results of pseudo-Hermitian uncertainty relation $R$. The quantum state is $|\psi\rangle=(\cos\theta_1\sin\theta_2, \cos\theta_1\cos\theta_2, \sin\theta_1)^{\text{T}}$ (without normalization) where $\theta_2$ is variable and $\theta_1$ is set as $0$ and $\frac{\pi}{2.5}$ marked in the figures. The uncertainty relation with the positive (non-positive) definite pseudo-Hermitian metrics are displayed in (a) and (b) ((c) and (d)). The dots with error bars are the experimental results. The solid lines depict the theoretical results. The dashed lines mark the limit $R=1$ of the pseudo-Hermitian uncertainty relation.}
\center
\end{figure}

\section*{Discussion}
Our study demonstrates the applicability and versatility of the non-Hermitian projective measurement method. Facilitated by the divisible direct-sum structure of our method, this measurement can be implemented by multiple projection respectively in each subspace for a given quantum state, making it suitable for applications in practical scenarios. Moreover, our non-Hermitian projective protocol is not limited in PH observables with real eigenvalues but can also be applied to other non-Hermitian systems containing complex eigenvalues, such as parity-time-symmetric broken systems \cite{Bender1998,Ozdemir2019}. Consider an inner product operation $\langle\psi_1,\psi_2\rangle$ of two states $|\psi_1\rangle$ and $|\psi_2\rangle$. The projective probability of a quantum state $|\psi\rangle$ onto a non-Hermitian eigenstate $|\Psi_k\rangle$ can always be expressed as $\text{Prob}(k)=\frac{\langle\Psi_k,\psi\rangle}{\langle\Psi_k,\Psi_k\rangle}$, provided that all the non-Hermitian eigenstates under this inner product satisfy two conditions: (1) mutual orthogonality and (2) non-vanishing norm. Furthermore, when focusing on the projection on non-Hermitian eigenstates, the required conditions to apply our non-Hermitian projective protocol based on direct-sum dilation are the aforementioned conditions (1) and (2), along with the requirement (3) that the non-Hermitian eigenstates are linearly independent, e.g., without coalescent eigenstates.

For an $n$-dimensional quantum system, in contrast to state tomography that requires counting on $n^2$ projective bases, our measurement approach requires the counting statistics on only $n$ projective bases. This reduction in measurement complexity is particularly advantageous in scenarios involving high-dimensional non-Hermitian dynamics with multiple and high-order exceptional points \cite{Zhong2018,Patil2022}. Our non-Hermitian measurement method allows focusing solely on eigenmode projection without the need for complete state information, offering a valid approach for directly measuring various non-Hermitian quantum systems. Furthermore, our experimental results demonstrate that the PH uncertainty relation indeed holds under the positive definite PH metric. It reveals that non-Hermitian systems could provide additional non-Hermitian observables for various quantum applications and investigations, such as quantum cryptography and quantum estimation \cite{Coles2017}.

The differences between non-Hermitian projective measurement and typical non-orthogonal generalized measurement, such as POVM, should also be discussed. In POVM, the measurement probability is $\text{Prob}(k)=\text{Tr}[\rho F_k]$, where $\{F_k\}$ are non-orthogonal Hermitian operators and $\sum_k F_k=\mathds{1}$ (noted the normalization coefficient is omitted for simplicity). In non-Hermitian measurement for PH observables, the measurement probability is given by $\text{Prob}(k)=\text{Tr}[\rho M_k]$, where $M_k=\frac{\eta|E_k\rangle\langle E_k|\eta}{\langle E_k|\eta|E_k\rangle}$ is also non-orthogonal Hermitian operator but $\sum_k M_k=\eta$ \cite{Supp}. The operator $\eta$ is usually not proportional to the identity operator, hence the operators $\{M_k\}$ cannot been transformed to the POVM measuring operators $\{F_k\}$ by the straightforward normalization. In the physical perspective, POVM can be interpreted as complete orthogonal measurement in a dilated high-dimensional space according to Naimark's theorem, which imposes restrictions on the number and structure of feasible POVM measuring operators. In contrast, non-Hermitian projective operators exhibit the specific skewed structure with the fixed number equaling to the system dimension, hence we utilize postselection in our dilated projective protocol to apply the effective non-Hermitian measurement.

In conclusion, we have investigated the projective measurement properties of PH observables and derive the PH uncertainty relation holding under positive definite PH metric. Next, we have designed a state-independent non-Hermitian projective measurement protocol utilizing direct-sum dilated projection combining with postselection. Subsequently, we experimentally constructed an non-Hermitian quantum simulator to implement the non-Hermitian quantum measurement for single-photon qutrits. We have experimentally obtained the measurement results of various 3-dimensional PH observables directly from the simulator output statistics, and experimentally investigated the PH uncertainty relation using the non-Hermitian quantum simulator. Our work presents and demonstrates an efficient approach for the direct quantum measurement of non-Hermitian systems, paving the route for exploring non-Hermitian quantum features from dynamics to measures.

\section*{Acknowledgements}

This work is supported by the Innovation Program for Quantum Science and Technology (No. 2021ZD0301200), the National Natural Science Foundation of China (Nos. 12174370, 12174376, 11821404 and 12304546), the Youth Innovation Promotion Association of Chinese Academy of Sciences (No. 2017492), the China Postdoctoral Science Foundation (No. 2023M733412), the Anhui Provincial Natural Science Foundation (No. 2308085QA28), and the Open Research Projects of Zhejiang Lab (No.2021MB0AB02).

\bibliographystyle{}

\begin{thebibliography}{29}
\bibitem{note1}
$\text{Tr}[P_k\rho_k]=\text{Tr}[\frac{1}{|\langle v_k|E_k\rangle|^2}|v_k\rangle\langle v_k|\rho]=\text{Tr}[\frac{1}{|\langle v_k|E_k\rangle|^2}|v_k\rangle\langle v_k|\sum_{ij}p_{ij}|E_i\rangle\langle E_j|]=\text{Tr}[\frac{1}{|\langle v_k|E_k\rangle|^2}|v_k\rangle\langle v_k|p_{kj}|E_k\rangle\langle E_j|]=p_{kk}$ where the trace is applied on the complete orthonormal bases containing $|v_k\rangle$.
\bibitem{note2}
Taking the detected single-photon counting projected in the $k$-th subspace as $m_k$ and its corresponding statistical weight as $w_k$, the statistical probability of the detected event in this subspace is derived by $\text{Prob}(k)=\frac{m_k}{\sum_k w_k m_k}$.
\end{thebibliography}

\end{document}